\documentclass{emulateapj}
\usepackage{natbib}
\usepackage{multirow}
\usepackage{graphicx}
\usepackage{subfloat}

\usepackage[usenames, dvipsnames]{color}

\shorttitle{}
\shortauthors{Zschaechner, et al.}

\begin{document}

\title{Spatially Resolved $^{12}$CO(2--1)/$^{12}$CO(1--0) in the Starburst Galaxy NGC\,253:  Assessing Optical Depth to Constrain the Molecular Mass Outflow Rate} 
\author{\sc Laura K. Zschaechner\altaffilmark{1, 2, 3}, Alberto D. Bolatto\altaffilmark{4}, Fabian Walter \altaffilmark{3}, Adam K.Leroy\altaffilmark{5, 6}, Cinthya Herrera\altaffilmark{7}, Nico Krieger\altaffilmark{3}, J.M.~Diederik Kruijssen\altaffilmark{8}, David S. Meier\altaffilmark{9, 10}, Elisabeth A. C. Mills\altaffilmark{11}, Juergen Ott\altaffilmark{10}, Sylvain  Veilleux.\altaffilmark{4}, and Axel Weiss\altaffilmark{12} }
\altaffiltext{1}{Finnish Center for Astronomy with ESO, FI-20014 Turun yliopisto, Finland; laura.zschaechner@helsinki.fi}
\altaffiltext{2}{University of Helsinki, P.O. Box 64, Gustaf H\"{a}llstr\"{o}min katu 2a, FI-00014 Helsingin yliopisto, Finland}
\altaffiltext{3}{Max Planck Institute f\"{u}r Astronomie, K\"{o}nigstuhl 17, 69117 Heidelberg, Germany}  
\altaffiltext{4}{Department of Astronomy and Joint Space Institute, University of Maryland, College Park, MD 20642, USA}
\altaffiltext{5}{Department of Astronomy, The Ohio State University, 140 West 18th Avenue, Columbus, OH 43210, USA}
\altaffiltext{6}{National Radio Astronomy Observatory, 520 Edgemont Road, Charlottesville, VA 22903, USA}
\altaffiltext{7}{Institut de Radioastronomie Millim\'{e}trique, 300 rue de la Piscine, Domaine Universitaire, 38406, Saint-Martin-d'H\'{eres}}
\altaffiltext{8}{Astronomisches Rechen-Institut, Zentrum f\"{u}r Astronomie der Universit\"{a}t Heidelberg, M\"{o}nchhofstraße 12-14, D-69120 Heidelberg, Germany}
\altaffiltext{9}{Department of Physics, New Mexico Institute of Mining and Technology, 801 Leroy Place, Socorro, NM 87801, USA}
\altaffiltext{10}{National Radio Astronomy Observatory - P.O. Box O, 1003 Lopezville Road, Socorro, NM 87801, USA}
\altaffiltext{11}{Department of Astronomy, Boston University, 725 Commonwealth Avenue, Boston, MA 02215, USA 0000-0001-8782-1992}
\altaffiltext{12}{Max-Planck-Institut f\"{u}r Radioastronomie, Auf dem H\"{u}gel 69, D-53121 Bonn, Germany}

\begin{abstract}

\par
   We present Atacama Large Millimeter/submillimeter Array (ALMA) observations of $^{12}$CO(1--0) and  $^{12}$CO(2--1) in the central 40\arcsec\ (680~pc) of the nuclear starburst galaxy NGC\,253, including its molecular outflow.   We measure the ratio of brightness temperature for CO(2--1)/CO(1--0), $r_{21}$, in the central starburst and outflow-related features. We discuss how $r_{21}$ can be used to constrain the optical depth of the CO emission, which impacts the inferred mass of the outflow and consequently the molecular mass outflow rate.  We find $r_{21}\lesssim1$ throughout, consistent with a majority of the CO emission being optically-thick in the outflow, as it is in the starburst. This suggests that the molecular outflow mass is $3-6$ times larger than the lower limit reported for optically thin CO emission from warm molecular gas.  The implied molecular mass outflow rate is $25-50$~M$_{\odot}$\,yr$^{-1}$, assuming that conversion factor for the outflowing gas is similar to our best estimates for the bulk of the starburst. This is a factor of $9-19$ times larger than the star formation rate in NGC\,253.   We see tentative evidence for an extended, diffuse CO(2--1) component.

\end{abstract}


\section{Introduction}\label{introduction}

\par
   Galactic-scale outflows/winds play a crucial role in galaxy evolution.  They have been theoretically predicted to suppress and quench star formation (e.g.\  \citealt{1986ApJ...303...39D}, \citealt{2013MNRAS.430.3213B}, \citealt{2015MNRAS.454.2691M}, \citealt{2016MNRAS.463.1431K}, \citealt{2017MNRAS.466.1213K}, \citealt{2017MNRAS.471..144S}), potentially enrich the intergalactic medium (IGM) \citep{2010MNRAS.406.2325O}, and may lead to the observed rarity of star-forming massive galaxies in the local universe (e.g.\ \citealt{2008MNRAS.391..481S}).  The importance of outflows stems from their mass loss, and the impact it has on the later evolution of the galaxy.  The mass outflow rate is especially important for the molecular phase, which has the potential to dominate the mass loss (e.g.\ \citealt{2015ApJ...814...83L}) and it is also closely related to star formation.  If molecular gas is ejected in large quantities, it will not be available to fuel future star formation. Only recently, with new millimeter and submillimeter arrays such as the Atacama Large Millimeter/submillimeter Array (ALMA), it has become feasible to study the molecular phase of these outflows in detail. 
   
\par
    Molecular gas is most easily traced using bright CO lines observed in the millimeter frequency regime.  A standard optical depth tracer is $^{12}$CO/$^{13}$CO.  However, $^{13}$CO is difficult to detect in relatively faint outflow features.    Alternatively, the $^{12}$CO(1--0) and $^{12}$CO(2--1) transitions\footnote{We use CO henceforth to refer to the main isotopologue $^{12}$CO} can be used to characterize the optical depth via the CO(2--1)/CO(1--0) brightness temperature ratio, $r_{21}$ (e.g.\ \citealt{1980ApJ...240...60K}, \citealt{1990ApJ...348..434E}).  While CO is generally optically-thick, it may be optically-thin in turbulent regions or in regions with high velocity dispersion in general such as winds or outflows.  Assuming the observed emission is from optically thick clouds that are warm (T$_{kin}\gtrsim20$\,K), and above the effective critical density ($n\gtrsim10^{2}-10^{3}$ depending on optical depth), $r_{21}$ should be approximately unity. There are, however, certain caveats to this approach, as $r_{21}$ is also dependent on temperature and density.  These degeneracies can be broken with constraints on the temperature and densities.  We assume temperatures consistent with appropriate estimates for conditions within the wind and utilize previous work by \citet{2017ApJ...835..265W} to provide density constraints.  These constraints are provided in $\S$~\ref{interpreting_r21}.

Because CO(2--1) has an inherently higher optical depth than CO(1--0),  in optically-thin gas this ratio has a theoretical limit of $r_{21}\approx4$ -- provided the gas is warm and its density is high enough to be in Local Thermodynamic Equilibrium (LTE).  Temperature gradients, density effects on excitation, or the presence of multiple gas components, however, can complicate this picture. Nonetheless, within the typical temperature range for molecular clouds the assumption that $r_{21}\sim$1 for optically thick emission holds to first order  (e.g.\ \citealt{1990ApJ...348..434E}, \citealt{2009AJ....137.4670L}). This lends the primary motivation for the work we present here: \textit{to determine whether or not the CO emission observed in the outflow of NGC\,253 is optically-thick}.  To assess the optical depth within the outflow, $r_{21}$ must be measured in any extended outflow features.

There are few high-resolution studies of molecular outflows.  Starburst-driven molecular outflows have been observed in a number of nearby galaxies, for example M 82 (\citealt{2002ApJ...580L..21W}, \citealt{2015ApJ...814...83L}), NGC\,253 \citep{2013Natur.499..450B,2017ApJ...835..265W}, NGC 1808 \citep{2016ApJ...823...68S}, and ESO320-G030 \citep{2016A&A...594A..81P}.  The mass-loss rates in these outflows were found to be comparable to the star formation rates of their host galaxies, although with uncertainties due to the geometry and mass estimation. In particular, there are large uncertainties associated with the molecular mass to CO luminosity ratio, especially in an environment that is very different from that of a galaxy disk (e.g.
 \citealt{2013ARA&A..51..207B}).  Obtaining accurate mass outflow rates is key to assess the full impact of galactic-scale molecular winds on the star formation processes within their host galaxies. 

At a distance of 3.5 Mpc \citep[{corresponding to $17.0~{\rm pc}$ per {\rm arcsec},}][]{2005MNRAS.361..330R}, with a systemic velocity of 243 km s$^{-1}$, NGC\,253 is the nearest nuclear starburst galaxy in the Southern sky, and it hosts a starburst-driven molecular outflow well-suited for detailed study.  This proximity, as well as the nearly edge-on orientation (inclination $i$ = 78$^{\circ}$) of NGC\,253, render it an ideal template for resolved studies of multiphase winds.  \citet{2013Natur.499..450B} found evidence for a prominent molecular wind, with a total molecular mass outflow rate of at least 3--9 M$_{\odot}$ yr$^{-1}$, amounting to 1--3 times the global star formation rate of NGC\,253. This estimate is a lower limit, because its calculation assumed optically thin CO emission from warm gas in the outflow.  
 
Not much is known about the optical depth of the CO emission in this environment. \citet{2015ApJ...801...63M} measure  CO(1--0)/C$^{17}$O(1--0) ratios of $\geq$350 in the starbursting inner disk of NGC~253, indicating a typical value of $\tau$ = 2--5 (for their assumed $^{16}$O/$^{17}$O isotopic ratio), corresponding to moderate optical depth.  Unfortunately, C$^{17}$O emission is much weaker than that of CO, so these measurements were not possible in the outflow.  \citet{2017ApJ...835..265W} find that some outflow features, such as the SW Streamer, have clear associated emission from high-dipole molecules (HCN, HCO$^+$, CS, and CN) commonly used to trace high-density molecular gas. The presence of these molecular species requires volume and surface densities that are hard to reconcile with optically thin CO emission, and suggest instead a larger mass for these features. There remains, however, the possibility that the high-dipole molecule emission is excited not by collisions with H$_2$, but by electron collisions, depending on the electron density in the outflowing gas \citep[see also][]{2017ApJ...841...25G}. Given a sufficiently high abundance of electrons, this mechanism could efficiently produce molecular emission from high-dipole molecules even in low density and low optical depth gas \citep[see discussion in ][]{2017ApJ...835..265W}.   In this work, we aim to assess the optical depth of the CO emission in order to further constrain the molecular mass of the outflow.

\par
   This paper is organized as follows:  $\S$~\ref{observations} presents the observations, data reduction, and initial presentation of the data.  Our analysis and results are presented in $\S$~\ref{results}, followed by a discussion in $\S$~\ref{discussion}.  The final results are summarized in $\S$~\ref{summary}.

\section{Observations \& Data Reduction}\label{observations}

\par
We use observations of the CO(1--0) and CO(2--1) line emission in NGC~253, obtained with ALMA using mosaicing mode.  CO (1--0) data are from ALMA Cycles 0 and 1 (program ID: ADS/JAO.ALMA\#2011.1.00172.S, 2012.1.00108.S; PI: Bolatto) with the 12-meter array and ALMA Compact Array (ACA), combined with Mopra observations to account for the short-spacings (PI:  J. Ott).   CO (2--1) data are from ALMA Cycle 2 (programme ID: ADS/JAO.ALMA\#2013.1.00191.S; PI: Bolatto) and include a 14-pointing mosaic with the 12-meter array using Nyquist sampling, ACA, and Total Power (TP) data.  The center of the observed region is offset from the center of NGC\,253 in order to include the full extent of the southern outflow.

\par
Observing dates, times, and calibrators pertaining to the CO(2--1) data presented here are listed in Table~\ref{tbl_1}.  Calibration and processing of the CO(1--0) is described in \citet{2015ApJ...801...25L}.  We use the same cube in this analysis and thus refer the reader to that paper for a full description of the CO(1--0) observations and data reduction.

\begin{deluxetable}{lr}
\tabletypesize{\scriptsize}
\tablecaption{Observational and Instrumental Parameters  \label{tbl_1}}
\tablewidth{0pt}
\tablehead
{
\colhead{Parameter} &
\colhead{Value}
}
\startdata
\phd \textbf{CO(1--0)} & see \citealt{2015ApJ...801...25L}\\

\\
\phd \textbf{CO(2--1)}\\
\phd Observation Dates$-$ACA &2014 Jun 28\\  
\phd Observation Dates$-$12m Array &2014 Dec 28\\
\phd Observation Dates$-$TP Array &2015 May 02 \\  
\phd \hspace{30pt} &2015 Jul 16\\
\phd \hspace{30pt} &2015 Aug 14\\
\phd \hspace{30pt} &2015 Aug 16\\
\phd \hspace{30pt} &2015 Aug 17\\
\phd Number of Antennas ACA&10\\
\phd Number of Antennas 12m Array&36\\
\phd Calibrators 12m \& ACA - Flux&Uranus\\
\phd \hspace{65pt} Gain&J0038-2459\\
\phd \hspace{65pt} Bandpass&J2258-2758\\
\phd Total Integration Time -- ACA&0.5 hrs\\
\phd Total Integration Time -- 12m Array& 0.35 hrs\\ 
\phd Total Integration Time -- TP Array& 6.75 hrs\\
\phd Calibrator TP&J0038+2459\\
\\

\phd Channel Spacing (final cubes) &5.0 km s$^{-1}$\\
\phd Beam Size (smoothed cubes)&1.9\arcsec$\times$1.4\arcsec\\ 
\phd \hspace{30pt} & 32$\times$24 pc\\
\phd RMS Noise (CO (1--0) smoothed cube)&2.5 mJy beam$^{-1}$\\ 
\phd \hspace{42pt} (CO (2--1) smoothed cube) &2.2 mJy beam$^{-1}$
\enddata

\end{deluxetable}

\par
    Calibration of the CO(2--1) data was done using the Common Astronomy Software Application (CASA).  Initial calibration used scripts provided by ALMA staff for the ACA data and the ALMA pipeline for the 12-meter data.  CASA version 4.2.2 was used for the pipeline and initial imaging.  The 12-meter and ACA data are of good quality and required minimal additional flagging.  A single iteration of phase-only self-calibration using line-free channels, followed by another iteration of phase and amplitude calibration of the 12-meter data improved image quality.  The same approach was taken for the self-calibration of the ACA data, but with the 12-meter data used as a model.  The 12-meter and ACA data were combined using {\tt CONCAT} followed by continuum subtraction using {\tt UVCONTSUB}.  Image cubes were created using Briggs weighting with a robust parameter of 0.5 in order to optimize both the sensitivity to diffuse gas and the resolution. To reduce sidelobes, a mask was created using the interactive mode of {\tt CLEAN}.  This mask was then used during a non-interactive {\tt CLEAN}.  A primary beam correction was applied to the cubes prior to combining them with TP data.  The cubes were then combined with the TP data using {\tt FEATHER} with a SD factor of 1 to recover missing flux (using CASA version 4.7.2).

The synthesized beam size of the original CO(2--1) cube is 1.7\arcsec$\times$1.0\arcsec with PA=80.9$^{\circ}$ and the rms noise is 2.0 mJy beam$^{-1}$ over a 5 km s$^{-1}$ channel. To maintain consistent spatial resolution for our analysis, the final cubes are smoothed to 1.9\arcsec$\times$1.4\arcsec, with a PA  of 78$^{\circ}$ in order to match the CO(1--0) cube.

\begin{figure*}
\begin{centering}
\includegraphics[width=180mm]{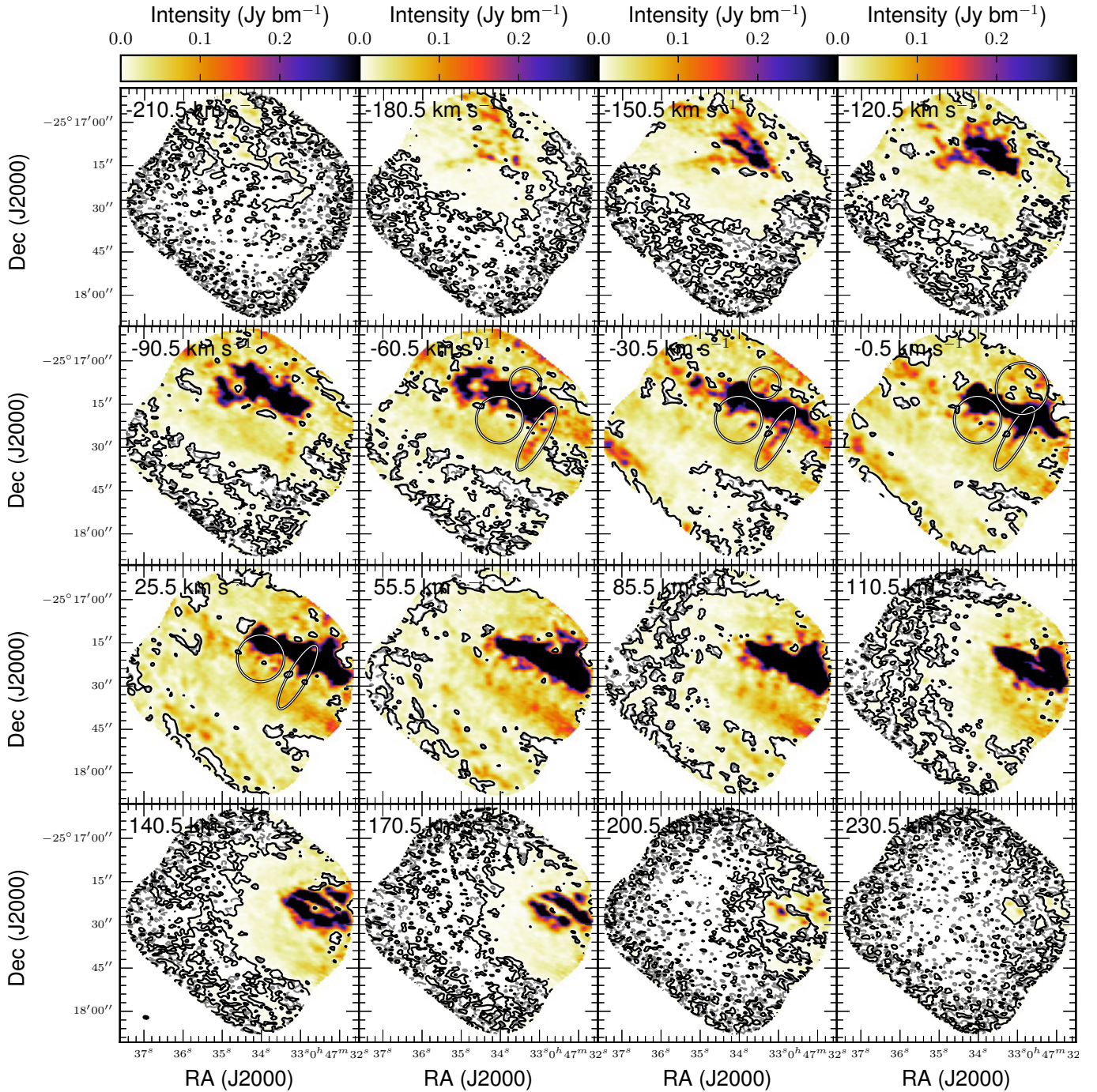}
\caption{\textit{Channel maps showing the CO(2--1) cube at the original (1.7\arcsec$\times$1.0\arcsec) resolution. Channels are integrated over 30 km s$^{-1}$ as opposed to the original km s$^{-1}$ widths. Velocities are given with respect to the systemic velocity.  The color scale is deliberately set to enhance both low-level outflow features and the $\sim$15$\sigma$ faint component extending $\geq$20\arcsec\ above the midplane. Black 3$\sigma$ and dashed gray --3$\sigma$ contours are shown.  The maximum signal is 3.6 Jy beam$^{-1}$, yielding a signal to noise ratio of 4400. The elongated ellipse marks the SW streamer \citep{2017ApJ...835..265W}, while the circles indicate select features related to the outflow, including emission at velocities not corresponding to rotation, and bubbles or shells. The emission to the southeast (bottom left, $-90.5$ to $+110.5$} km\,s$^{-1}$) is from a spiral arm.  The beam is shown in the bottom left panel.  The CO(1--0) cube is presented in \citet{2015ApJ...814...83L}.  \label{channel_maps}}
\end{centering}
\end{figure*}

\section{Results and Analysis}
\label{results}

\begin{figure}
\includegraphics[width=80mm]{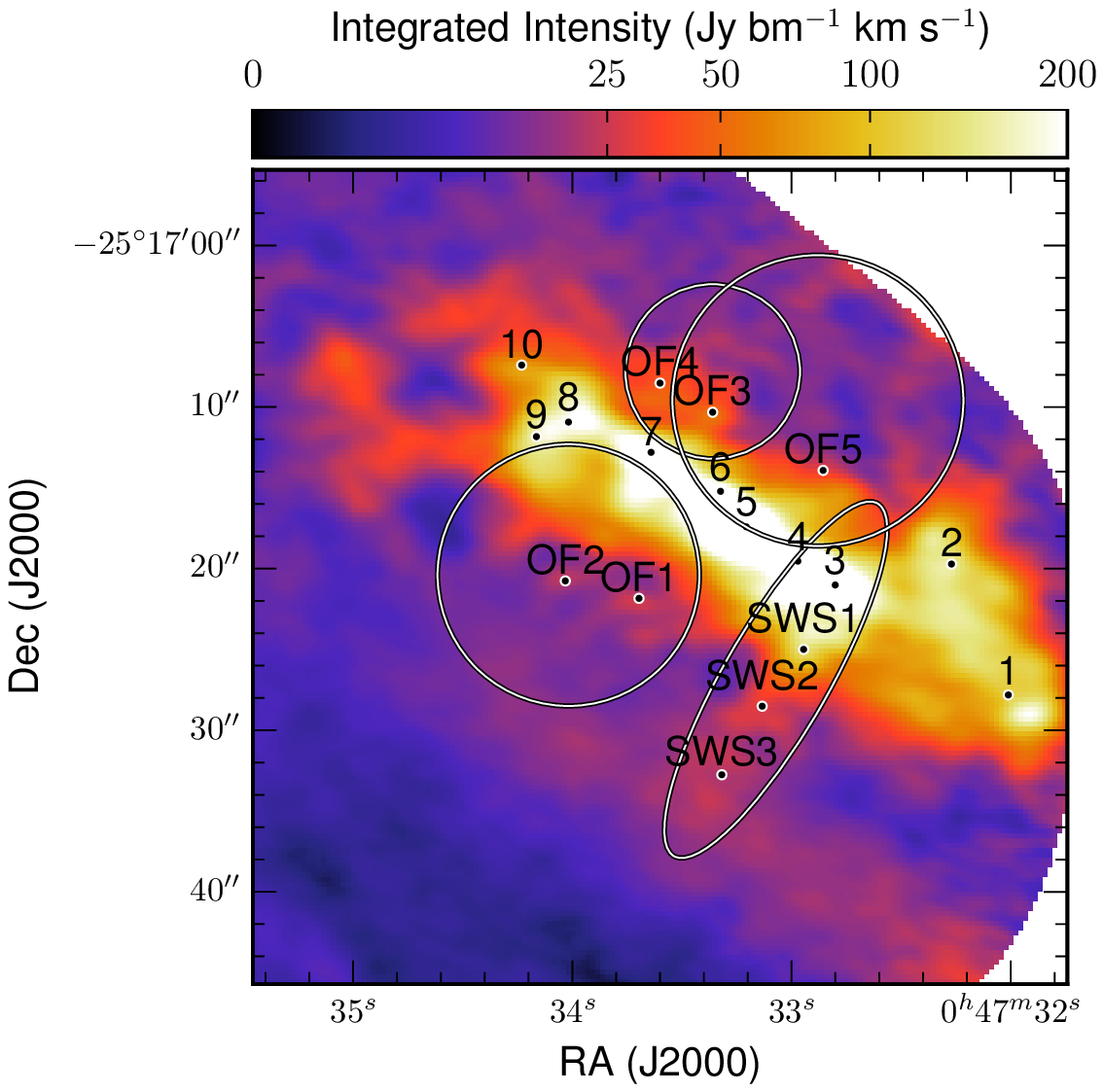}
\caption{\textit{An integrated intensity map of the CO(2--1) at a 1.9\arcsec$\times$1.4\arcsec\ resolution.  The map is created integrating all emission over a 2$\sigma$ threshold (4.4 mJy beam$^{-1}$ in a 5 km s$^{-1}$ channel), for channels from --208 to 222 km s$^{-1}$.  The ellipses are the same as in Figure~\ref{channel_maps}.  The labeled points correspond to the locations of spectra shown in Figures~\ref{spectra_grid_LeroyMatch_no32},~\ref{spectra_grid_SouthernStreamer_no32}, and~\ref{spectra_grid_OF_no32}. } \label{spectra_map_LeroyMatch}}
\end{figure}

We show the full resolution CO(2--1) data in Figure~\ref{channel_maps}  There is emission from an extended component that can be seen at a typical level of $\geq$30$\sigma$ ($\sim$30 mJy beam$^{-1}$ over 30 km s$^{-1}$, Figure~\ref{channel_maps}) in all channels between --180.5 and 170.5 km s$^{-1}$ km s$^{-1}$ with respect to the systemic velocity of 243 km s$^{-1}$.  We discuss the reality of this emission in $\S$~\ref{flux_recovery}.  Figure~\ref{spectra_map_LeroyMatch} shows a zeroth-moment map of the CO(2--1) created with {\tt IMMOMENTS} and including all emission above a 2$\sigma$ threshold in channels ranging from --208 to 222 km s$^{-1}$ at 5 km s$^{-1}$ velocity resolution.  
Prominent outflow features can be seen throughout the data, with the most notable being the so-called SW streamer \citep{2017ApJ...835..265W} from -60.5 km s$^{-1}$ to 25.5 km s$^{-1}$ (highlighted by the elongated ellipse in Figure~\ref{channel_maps}).  This feature and its comparison to the main disk are the primary focus of the analysis presented in this section.

\subsection{Obtaining $r_{21}$}
\par
To simplify the interpretation of the line ratios in physical terms, we convert the observations from flux density to surface brightness, expressed in Rayleigh-Jeans temperature units, and perform the correction for the Cosmic Microwave Background (see, for example, Eq. 6 in \citealt{2013ARA&A..51..207B}).  To obtain $r_{21}$, we divide the CO(2--1) cube by a similarly-corrected CO(1--0) cube.   

\par
 In order to avoid division by zero, the thresholds are initially set at 20 mJy\,beam$^{-1}$ in the CO(1--0) cube.  After taking the ratio of the CO(2--1) and CO(1--0) cubes, artificially high ratios on the edges are eliminated by creating a mask, which is then smoothed with a Gaussian using {\tt IMSMOOTH} to a target resolution of 2.0$\arcsec$$\times$$1.5\arcsec$. (The smoothing is elliptical as opposed to circular in order to preserve resolution of the 1.9$\arcsec$$\times$$1.4\arcsec$ beam of the cubes.  The orientation of the smoothing Gaussian is the same as that of the cubes.)  Emission that falls within 70$\%$ of the peak of the smoothed mask is then included in the final ratio cube. This threshold is somewhat arbitrary, but is chosen because it excludes the border artifacts while including as much emission associated with narrow outflow features as possible.  We then apply this final mask to the ratio cube, which effectively eliminates any artificially-high border values.

\begin{figure*}
\includegraphics[width=180mm]{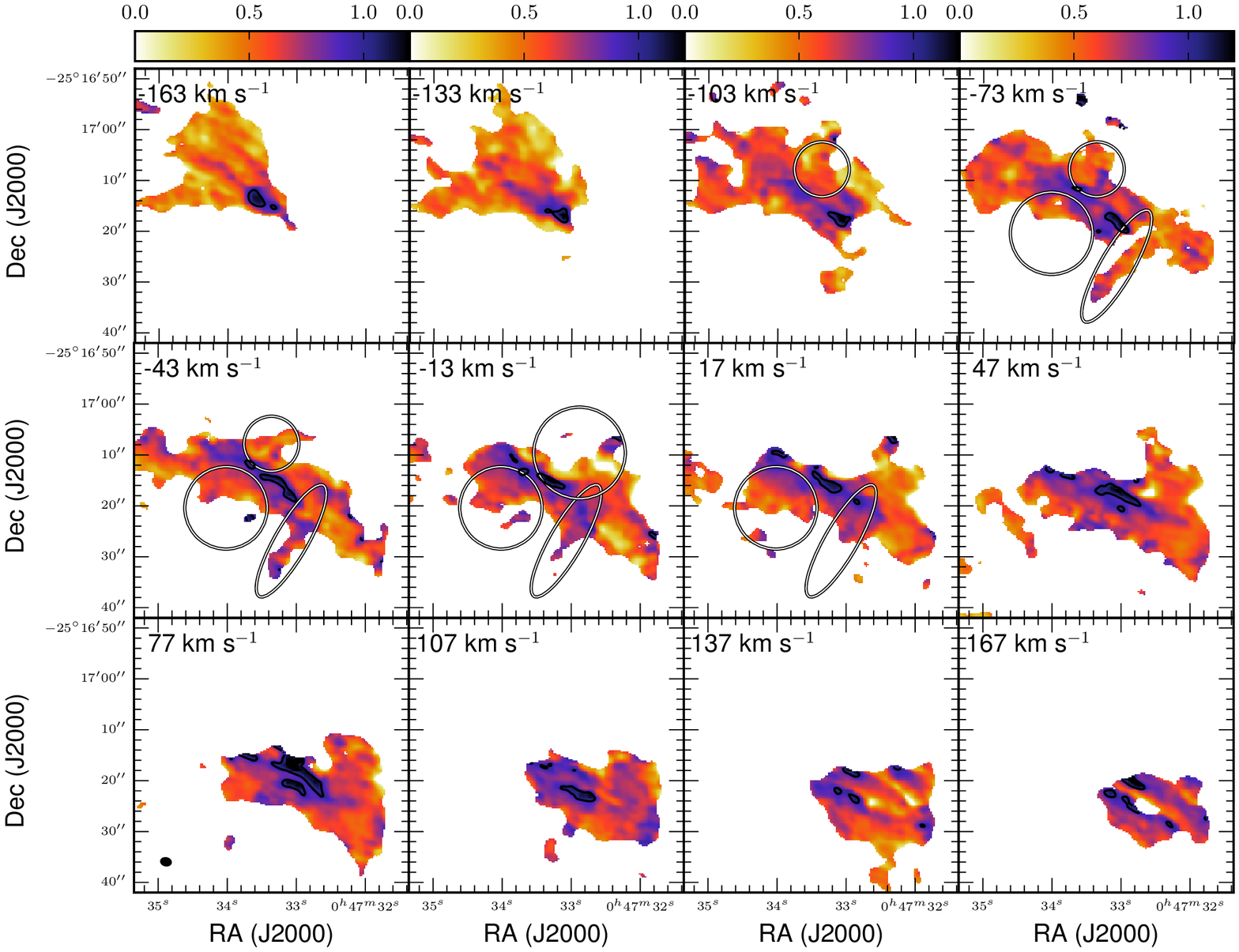}
\caption{\textit{$r_{21}$ ratio maps of NGC\,253.  The masking procedure described at the beginning of $\S$~\ref{results} has been applied.   The black contour is unity.  Annotations are as in Figure~\ref{channel_maps}.} \label{channel_maps_ratio_21}}
\end{figure*}

\subsection{Interpreting $r_{21}$}
\label{interpreting_r21}
There are a number of treatments of the basics of CO line ratios in the literature (see, for example, \citealt{1990ApJ...348..434E}), but since this is at the core of the measurement we summarize the main equations and results here. The observed flux density for a $J\rightarrow J-1$ transition at frequency $\nu$,  $S^{obs}_\nu$, is

\begin{equation}
S^{obs}_\nu=(1-e^{-\tau_J})(B_\nu(T_{ex,J})-B_\nu(T_{cmb})),
\end{equation}

\noindent where $T_{ex,J}$ is the excitation temperature of the transition, $T_{cmb}$ is the temperature of the Cosmic Microwave Background (CMB), $B_\nu$ is the Planck function, and $\tau_J$ is the optical depth of the transition. From this we can compute that the observed Rayleigh-Jeans brightness temperature for the same transition, which for simplicity we will note below by its upper quantum number as $T_J$, is 
    
\begin{equation}
T_J=(1-e^{-\tau_J})\frac{h\nu_J}{k}\left({\frac{1}{e^\frac{h\nu_J}{kT_{ex,J}}-1}-\frac{1}{e^\frac{h\nu_J}{kT_{cmb}}-1}}\right),
\end{equation}\label{T_J}

\noindent where $\nu_J$ is the frequency and the excitation temperature of the transition, and $h$ and $k$ are the Planck and Boltzmann constants. For rotational transitions $\nu_J=J\nu_1$. The CMB correction is usually a small effect, that in terms of Rayleigh-Jeans brightness temperature amounts to $0.84$ and $0.20$~K at the frequencies of the $1-0$ and $2-1$ transitions respectively.

If we ignore the small CMB correction in order to simplify the equations, the ratio of brightness temperatures for two consecutive transitions is

\begin{equation}
\frac{T_{J+1}}{T_J}\approx\left(\frac{1-e^{-\tau_{J+1}}}{1-e^{-\tau_J}}\right)\left(\frac{J+1}{J}\right)\left(\frac{e^\frac{Jh\nu_1}{k T_{ex,J}}-1}{e^\frac{(J+1)h\nu_1}{k T_{ex,J+1}}-1}\right).\label{eq:3}
\end{equation}

\noindent For the low $J$ transitions of CO and because of their low effective critical densities we can assume LTE, that is, the distribution of the population among energy levels will follow the Boltzmann equation at the physical temperature of the system. In that case, their excitation temperature will be simply the kinetic temperature of the gas $T_{ex,J}=T_{ex,J+1}=T_{kin}$ (a somewhat more general assumption, thermalization, implies $T_{ex,J}=T_{ex,J+1}$ and is equivalent for the calculations below). Attaining LTE requires densities to be large, in principle larger than the critical density of the highest transition: the critical density for CO(1--0) is $n_{cr}\approx2.2\times10^3$~cm$^{-3}$ at $T\sim30~K$ \citep{2010ApJ...718.1062Y}, with a weak dependency on temperature and a strong $\propto J^3$ dependency on the upper level of the transition. For optically thick lines, however, radiative trapping increases the excitation resulting in lower effective critical densities $\sim n_{cr}/\tau$. This implies that, in gas with moderate optical depth, LTE-like conditions can be reached for $n\sim 10^3$~cm$^{-3}$. Moreover, because $h\nu_1/k=5.53$~K, for emission arising from warm gas $T_{kin}\gg h\nu/k$ for the low $J$ transitions of CO. In those conditions the last factor in Eq. \ref{eq:3} can be simplified to be $\sim J/(J+1)$, canceling the central factor. Therefore, we expect $r_{21}$ ratios approaching unity for optically thick ($\tau\gg1$) emission arising in warm ($T_{kin}>11$~K), moderately dense ($n>10^3$~cm$^{-3}$) gas.

In the case of optically thin emission ($\tau\ll1$), on the other hand, the first factor in Eq. \ref{eq:3} simplifies to be $\sim\tau_{J+1}/\tau_J$. The optical depth for the transition $J\rightarrow J-1$ is simply

\begin{equation}
\tau_J=\frac{8\pi^3\mu^2}{3h}\frac{J}{2J+1}\left({e^\frac{Jh\nu_1}{kT_{ex,J}}-1}\right)\frac{N_J}{\Delta v},
\end{equation}

\noindent where $\mu$ is the dipole moment, $N_J$ is the column density in the upper level of the transition, and $\Delta v$ is the velocity dispersion. Thus, if the two transitions arise from the same parcel of gas (and therefore have the same velocity dispersion), Eq. \ref{eq:3} is reduced to

\begin{equation}
\frac{T_{J+1}}{T_J}\approx\left(\frac{J+1}{J}\right)^2 e^{-\frac{(J+1)h\nu_1}{k T_{ex,J+1}}},
\end{equation}

\noindent implying that in warm gas, $r_{21}$ can be as high as $4$. Note that the ratio of optical depths for thermalized emission from ``warm'' gas is simply

\begin{equation}
\frac{\tau_{J+1}}{\tau_J}\approx\left(\frac{J+1}{J}\right)^2, 
\label{tauratio}
\end{equation}

\noindent implying that the optical depth grows faster for the higher transition than for the lower one.

In the presence of temperature gradients, the $r_{21}$ ratio may be somewhat larger than unity even for optically thick emission, but it is extremely unlikely to reach as high a value as in warm, optically-thin gas.  Ratios less than unity, $r_{21}<1$, can also be caused by temperature gradients, or alternatively for subthermal excitation (i.e., $T_{ex}<T_{kin}$). The latter is very unlikey in NGC~253, particularly in its central starbursting regions.  Because of Eq. \ref{tauratio}, optically thick emission arising from regions with temperature gradients where the temperature is increasing into the emitting cloud, are a more likely explanation for $r_{21}<1$.

\iffalse
\begin{subfigures}
\begin{figure*}
\includegraphics[width=1\linewidth,bb=0 0 200 200]{spectra_grid_LeroyMatch_no32.pdf}
        \caption{}\label{spectra_grid_LeroyMatch_no32}
\end{figure*}
\begin{figure*}
\includegraphics[width=1\linewidth,bb=0 0 150 200]{spectra_grid_SouthernStreamer_no32.pdf}
        \caption{}\label{spectra_grid_SouthernStreamer_no32}
\end{figure*}

\begin{figure*}
\includegraphics[width=1\linewidth,bb=0 0 150 200]{spectra_grid_OF_no32.pdf}
\caption{\textit{CMB corrected T$_{J}$, excluding the negligible optical depth term (Eq.~\ref{T_J}), extracted from cubes corresponding to points 1--10 from Figure~\ref{spectra_map_LeroyMatch} are shown in ``A", while spectra corresponding to the SW streamer are shown in ``B" and other potential outflow features are shown in ``C".   Solid lines represent spectra that include short-spacing corrections from either the Mopra or the TP array.  Dashed lines represent the same spectra, but without short-spacing corrections. } \label{spectra_grid_OF_no32}}
\end{figure*}
\end{subfigures}
\fi

\begin{figure*}
\includegraphics[width=1\linewidth,bb=0 0 200 200]{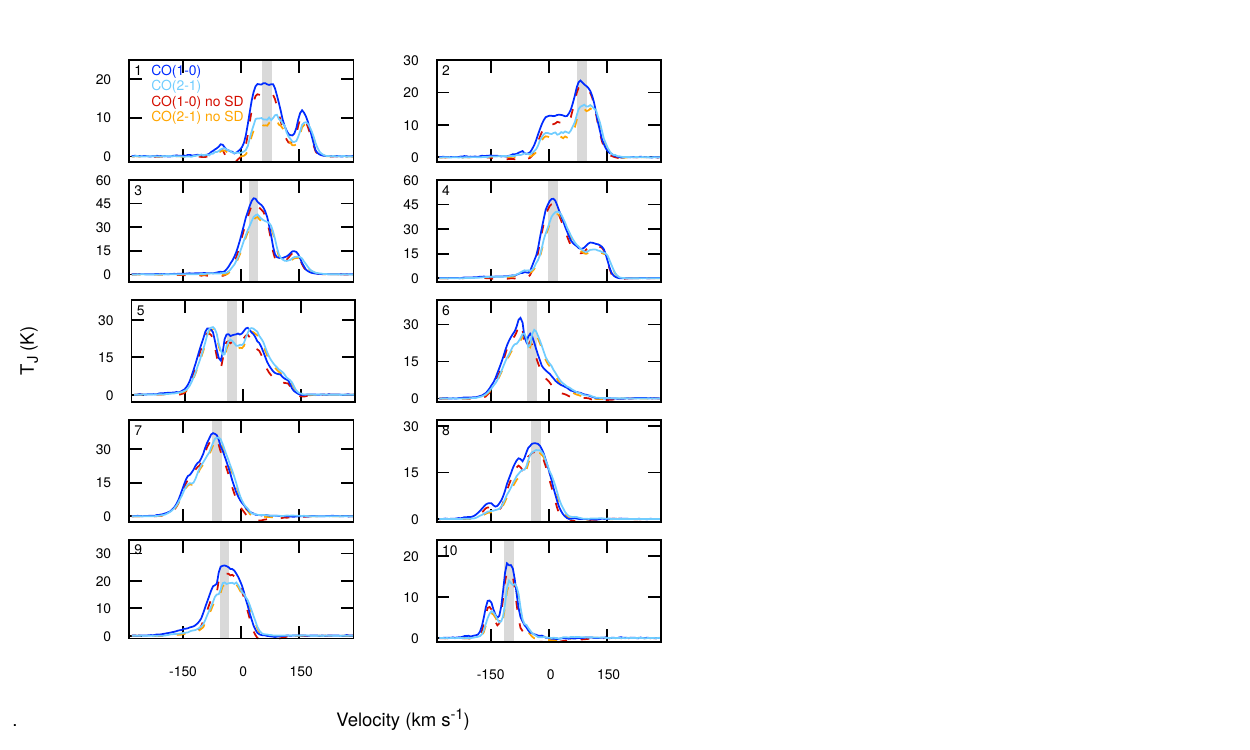}
        \caption{\textit{Spectra extracted from unmasked cubes corresponding to points 1--10 from Figure~\ref{spectra_map_LeroyMatch}.  T$_{J}$ is the Rayleigh-Jeans brightness temperature as defined at the beginning of $\S$~\ref{results}.  Solid lines represent spectra that include short-spacing corrections from either the Mopra or the TP array.  Dashed lines represent the same spectra, but without short-spacing corrections. The shaded gray regions represent the velocity range over which r$_{21}$ measurements are taken.}}\label{spectra_grid_LeroyMatch_no32}
\end{figure*}

\begin{figure*}

\includegraphics[width=1\linewidth,bb=0 0 160 200]{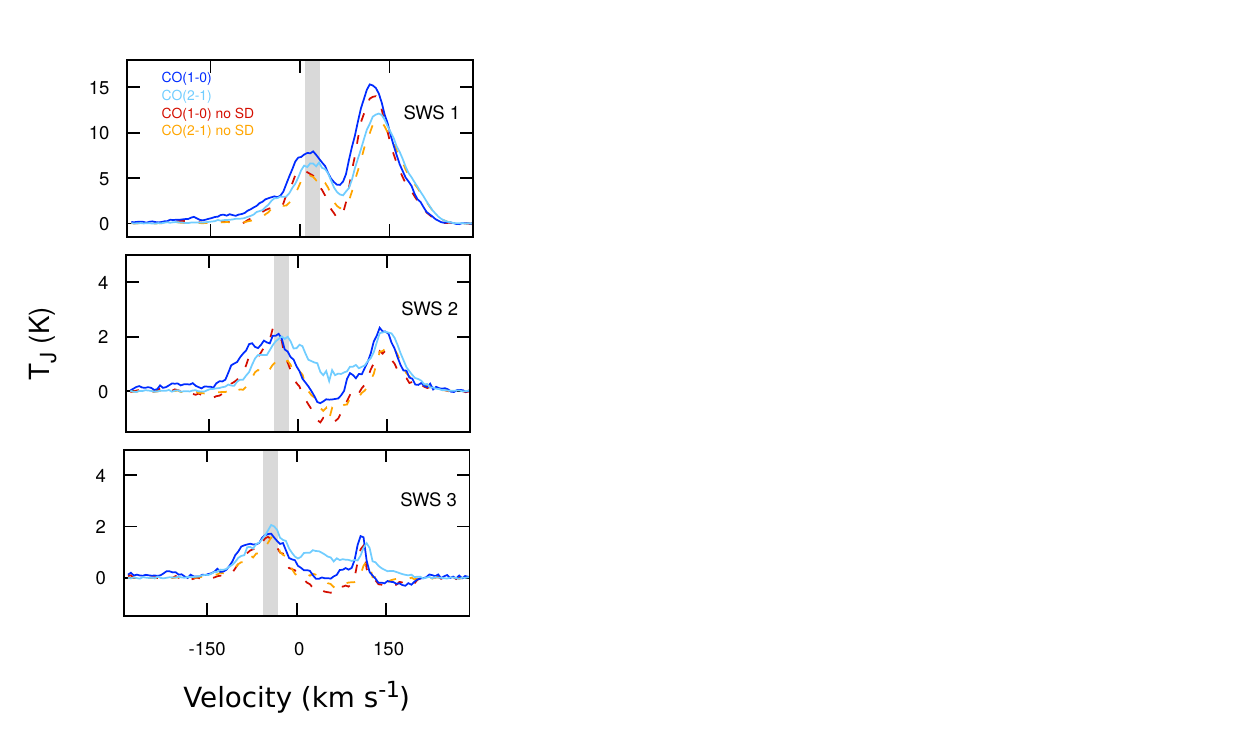}
        \caption{\textit{Same as Figure~\ref{spectra_grid_LeroyMatch_no32} but for spectra extracted from unmasked cubes corresponding to the SW streamer (labeled as ``SWS" in Figure~\ref{spectra_map_LeroyMatch}  }}\label{spectra_grid_SouthernStreamer_no32}
\end{figure*}

\begin{figure*}

\includegraphics[width=1\linewidth,bb=0 0 160 200]{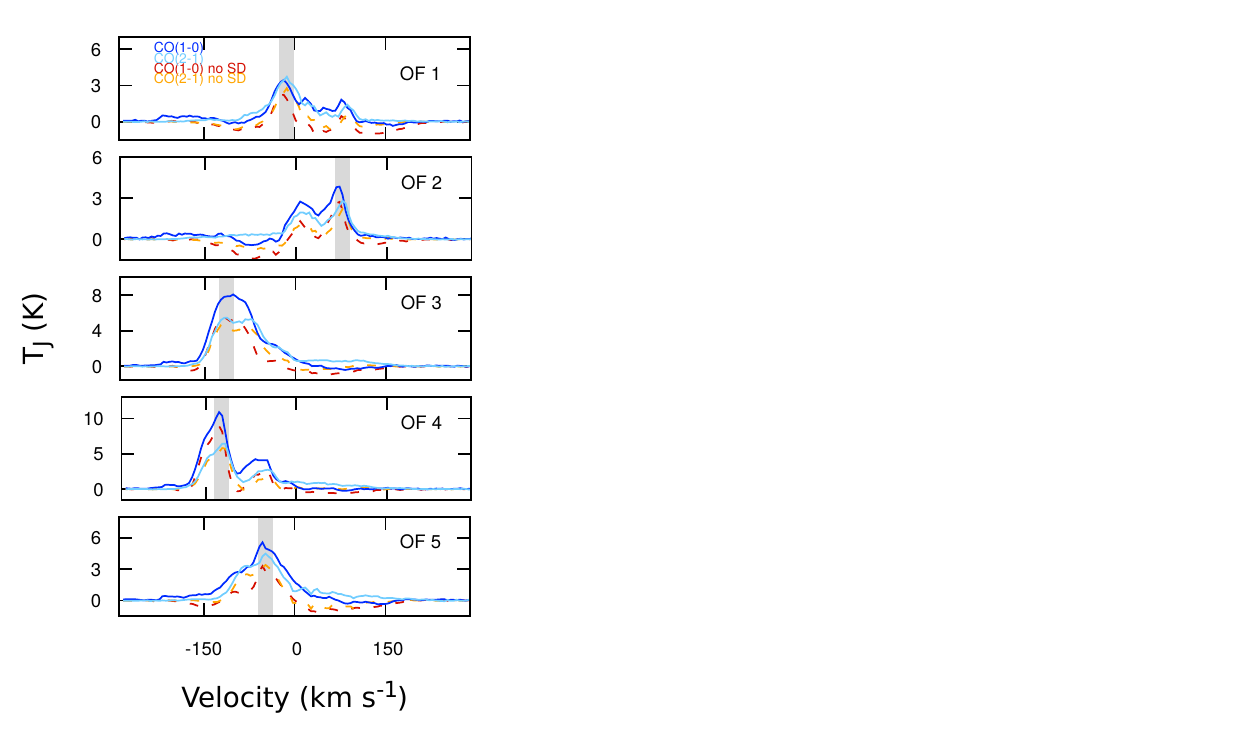}
\caption{\textit{Same as Figure~\ref{spectra_grid_LeroyMatch_no32} but for additional potential outflow features labeled as ``OF" in Figure~\ref{spectra_map_LeroyMatch}.} \label{spectra_grid_OF_no32}}
\end{figure*}

\begin{deluxetable*}{lccccccc}
\tabletypesize{\scriptsize}
\tablecaption{Molecular Cloud Locations From \citet{2015ApJ...801...25L} and Additional Outflow Features  \label{tbl_2}}
\tablewidth{0pt}
\tablehead
{
\colhead{Region Number} &
\colhead{R.A.}&
\colhead{Decl.}&
\colhead{$r_{21}$}\tablenotemark{a}&
\colhead{2$\sigma$}\tablenotemark{b}&
\colhead{$r_{21}$*}\tablenotemark{c}&
\colhead{$\%$ difference}\tablenotemark{d}&
}
\startdata
\phd 1&00 47 32.01&--25 17 27.8&0.51\tablenotemark{e}&0.04&0.51&1$\%$\\   
\phd 2&00 47 32.27&--25 17 19.7&0.67\tablenotemark{e}&0.04&0.66&2$\%$\\
\phd 3&00 47 32.80&--25 17 21.0&0.82&0.06&0.84&3$\%$\\
\phd 4&00 47 32.97&--25 17 19.5&0.83&0.07&0.86&3$\%$\\
\phd 5&00 47 33.21&--25 17 17.4&1.01&0.12&1.05&4$\%$\\
\phd 6&00 47 33.32&--25 17 15.2&0.95&0.19&1.04&9$\%$\\
\phd 7&00 47 33.64&--25 17 12.8&0.93&0.08&0.95&3$\%$\\
\phd 8&00 47 34.02&--25 17 10.9&0.85&0.03&0.92&7$\%$\\
\phd 9&00 47 34.16&--25 17 11.8&0.72&0.08&0.79&6$\%$\\
\phd 10&00 47 34.23&--25 17 07.4&0.75&0.09&0.81&7$\%$\\
\phd SW streamer 1&00 47 32.95&--25 17 25.0&0.79&0.03&0.84&6$\%$\\
\phd SW streamer 2&00 47 33.13&--25 17 28.5&0.78&0.12&0.49&46$\%$\\
\phd SW streamer 3&00 47 33.32&--25 17 32.8&0.83&0.10&0.67&22$\%$\\
\phd Outflow 1 &00 47 33.69&--25 17 21.8&0.91&0.12&0.95&5$\%$\\
\phd Outflow 2 &00 47 34.03&--25 17 20.6&0.60&0.05&0.62&2$\%$\\ 
\phd Outflow 3 &00 47 33.36&--25 17 10.3&0.65\tablenotemark{e}&0.10&0.77&18$\%$\\
\phd Outflow 4 &00 47 33.60&--25 17 08.5&0.55\tablenotemark{e}&0.50&0.61&9$\%$\\ 
\phd Outflow 5 &00 47 32.86&--25 17 13.9&0.70\tablenotemark{e}&0.07&0.89&24$\%$
\enddata
\tablenotetext{a}{Average value for the ratio in each region for the range of channels specified in the text.  Values include CMB correction.}
\tablenotetext{b}{2$\sigma$ values over the measured velocity range.}
\tablenotetext{c}{Same as (a), but for cubes without short-spacing corrections.}  
\tablenotetext{d}{The percent difference between $r_{21}$ and $r_{21}$* defined as 100$\times$$\vert$$r_{21}-r_{21}$*$\vert$/$\frac{1}{2}$($r_{21}$+$r_{21}$*).}
\tablenotetext{e}{Values of $r_{21}$ are likely artificially low due to missing flux at the edge of the map.
}

\end{deluxetable*}

\subsection{Measurements in NGC\,253}

\par
We measure $r_{21}$ in selected regions in NGC\,253 over a 1.$\arcsec$9$\times$1.$\arcsec$4 aperture in all cases. We make this measurement in the regions marked 1--10 from \citet{2015ApJ...811...15L}.  We also measure $r_{21}$ along the SW streamer (labeled as SWS 1--3) and in five additional features with kinematics and morphology consistent with outflowing gas (labeled OF 1--5).  The locations of these figures are shown in Figure~\ref{spectra_map_LeroyMatch}.

\par
We are particularly interested in obtaining a measure of $r_{21}$ at the velocities where the emission peaks. This maximizes the SNR of the measurement, and focuses it on the spectral range that contributes the most to the emission. Thus, for the selected regions, we measure the ratio of CO(1--0) and CO(2--1) brightness temperatures over custom 30 km s$^{-1}$ velocity ranges.  These ranges encompass both the CO(1--0) and CO(2--1) peak values in most regions. There are some exceptions, outlined below.   These spectral ranges are shown as shaded gray regions in Figure~\ref{spectra_grid_LeroyMatch_no32}--~\ref{spectra_grid_OF_no32}.

The peaks in regions 5 and 6 are less distinct than in other regions.  Furthermore, both of these regions show some variation in r$_{21}$.  For region 5, we have selected the velocity range over which to measure r$_{21}$ to fall within the approximate center of the flattened broader emission peak.  For region 6, we select a spectral range that contains the peak of the CO(2--1) emission, as well as a local maximum for the CO(1--0) emission.

In the SW streamer there are two distinct spectral features: one (toward negative velocities) is associated with the outflowing gas, while the other (toward positive velocities) corresponds to gas that is in the background of the streamer, participating in the rotation of the disk of NGC\,253. Our measurements of $r_{21}$ are in the spectral feature corresponding to the outflowing gas.

Using this procedure, we find that the $r_{21}$ ratio takes values between 0.5 and 1.0. In particular, we do not measure values of $r_{21}$ that would be indicative of optically thin emission in the SW streamer. We present these values in Table~\ref{tbl_2}, and discuss the effects of spatial filtering in $\S$~\ref{flux_recovery}.  

The ratio $r_{21}$ is significantly less than unity in regions corresponding to regions 1 and 2.  The CO(2--1) and CO(1--0) spectra in these regions are very similar, but have different amplitudes (Figure  \ref{spectra_grid_LeroyMatch_no32}).  Their similar kinematics indicate that the emission from each is likely co-spatial.  Pointing 2 is near the base of an outflow bubble that has burst in the NW direction (best seen at 55.5 km s$^{-1}$ in Figure~\ref{channel_maps} near the NW edge of the map and in the faint emission of Figure~\ref{spectra_map_LeroyMatch}).  At 110.5 km s$^{-1}$ in Figure~\ref{channel_maps}, it can be seen that pointing 1 is near a relatively faint, diffuse feature that extends above the disk such that it is parallel to the SW Streamer, although not coincident.

\subsection{Flux Recovery}\label{flux_recovery}

The degree to which we recover extended emission can have an important impact on the measured line ratios. Our interferometric maps include ACA and single dish, so we expect them to do a good job at recovering emission on all spatial scales. In order to assess how well we are doing, 
we compare the ALMA data with single-dish observations in the literature.  We smooth the cubes using {\tt IMSMOOTH} to a 23\arcsec\ FWHM beam size in order to match the resolution presented in \citet{1996A&A...305..421M} and \citet{1999MNRAS.303..157H}.   After smoothing, we convert from mJy beam$^{-1}$ to K. We then create integrated intensity maps over the full velocity range of the emission  $-$201--229 km s$^{-1}$ and compare the integrated fluxes with those provided in \citet{1999MNRAS.303..157H} over the same region. Our total CO(1--0) flux is 650 K km s$^{-1}$.  This value is $\sim$72$\%$ of the single dish value presented in \citet{1996A&A...305..421M} using the Institut de Radioastronomie Millim\'{e}trique (IRAM) 30-meter telescope.  
The flux we measure for CO(2--1) is 840 K km s$^{-1}$, which is 79$\%$ of the flux listed in \citet{1999MNRAS.303..157H}, derived from James Clerk Maxwell Telescope (JCMT) observations, and is slightly higher than the value reported by \citet{1996A&A...305..421M} using the IRAM 30-meter telescope. Based on this we conclude that our fluxes measured within the same 23\arcsec\ aperture and at 23\arcsec\ resolution are very comparable to previous measurements of NGC\,253, since uncertainties at the level of 20\% and larger are not uncommon.

To further assess the impact of the uncertainties associated with the combination of single-dish and interferometer data on $r_{21}$, we compare line ratios obtained from maps that have single-dish total power information included, and maps that do not have it. The results are presented in the sixth column of Table~\ref{tbl_2}. The absence of short-spacing corrections typically affect the ratios by $<$10\%, although the effect can be as large as 50\% in the most extreme case. Note that including the short-spacing correction in principle improves the accuracy of the flux measurement, therefore the $r_{21}$ values that include them in the fourth column should be the ones used.

Imperfect recovery of the large spatial scales usually manifests itself in a number of artifacts in the map, particularly the presence of a negative depression (a ``bowl'') surrounding the bright emission that is due to the attenuation of the large spatial scales. Similar artifacts can also be present due to imperfect deconvolution (cleaning) of the interferometric images. As can be seen in Figure~\ref{spectra_grid_LeroyMatch_no32}, regions 1--10 do not show substantial negative ranges in their spectra, suggesting that this is not affecting the values of $r_{21}$ measured for the bright emission. This can be more of a concern for measurements of $r_{21}$ in faint emission near bright emission.  
Regions in the SW streamer (Figure~\ref{spectra_grid_SouthernStreamer_no32}) show some velocity ranges with negative values. Note, however, that this tends to affect mostly the CO (1--0) spectra. However, these velocity ranges are not included in our $r_{21}$ measurements.
  
It should be noted that some of the regions are near the edge of the map.  Consequently, they may suffer from poor flux recovery.  In particular, this issue likely affects regions 1 and 2, as well as OF 3--5 in Table~\ref{tbl_1}.  In these cases $r_{21}$ could be artificially low.

\subsection{The Extended CO(2--1) Component}\label{extended_component_test}
\par
As Figure~\ref{channel_maps} shows, there is widespread faint, extended emission on scales of $\sim20\arcsec$ (340~pc) surrounding the bright regions, with typical intensities of  ($\sim$20--85 mJy beam$^{-1}$. This component is mostly due to the combination of the total power (single-dish) data with the interferometric observations. 

The question is whether this widespread emission is real, or a feature introduced in the combination: given that the peak flux of the central regions of NGC\,253 is a few Janskys per beam, the extended emission could be due to imperfections in the single-dish map at the level of a few percent, or possibly an error in the relative amplitude calibration of the single-dish and interferometric data. We have tried to reduce the chance of introducing artifacts when combining single-dish and interferometric data.  We started by flagging $\sim$20\% of the total-power data showing possible baseline problems that may require high-order polynomial fits.  Following flagging, we did the baseline removal by fitting channels free of emission (i.e.\ channels with absolute velocities between --300 to --10 km s$^{-1}$ and from 500 to 700 km s$^{-1}$), with a first order polynomial.  Proceeding in this manner removed a fraction of the extended component present in our original combination using the total-power data as delivered, but did not remove all the extended emission.   The remaining extended emission is roughly contained between approximately $-$100 and 170 km s$^{-1}$ with respect to the systemic velocity.  It is possible that there is a relative amplitude calibration error, but it would have to be large to explain the observation.

In terms of its impact on the $r_{21}$ values we discuss, the extended component is relatively unimportant. It has a peak brightness of 150 mJy beam$^{-1}$, with typical values of 40 mJy beam$^{-1}$, and is spread throughout a 40\arcsec\ region surrounding the main disk. In most cases, because we measure $r_{21}$ over a narrow spectral range including the emission peaks, the extended component contributes at or less than a few percent to the CO(2--1) value, and consequently, a negligible amount to $r_{21}$. 

Ultimately, we are unable to demonstrate that the extended emission is an artifact.  We thus tentatively propose that it is a real feature, with some degree of skepticism, and discuss its properties further in $\S$~\ref{haze}.  

We see no indication of an analogous component in the existing CO(1--0) data.  However, the imperfect zero-spacing correction, as indicated by the slight negatives surrounding the brightest emission in those data, may make the detection of an extended component difficult.  Where present, the negative values are typically 4--6 $\sigma$ (0.2--0.3 K) in a 5 km s$^{-1}$ channel, while the CO(2--1) extended component is $\sim$0.4 K.  Given that these are worse case scenarios for the detection of an extended component, if an extended CO(1--0) component exist it has to be considerably fainter than that observed in CO(2--1). Thus, the corresponding r$_{21}$ values would be high, indicating a low optical depth for the extended component.

\section{Discussion}\label{discussion}

\subsection{Optical Depth in the Starburst and Outflow}
\label{optical_depth}

The ratios in Table \ref{tbl_2} are most compatible with the hypothesis that the CO emission is optically thick in both the starburst and the different outflow regions. For the starburst region this is not surprising. Indeed, \citet{2015ApJ...801...63M} do a more direct test of the optical depth of CO in this region by comparing the $J=1\rightarrow0$ emission from the C$^{18}$O and C$^{17}$O isotopologues, and conclude that the CO(1--0) transition has $\tau\sim2-6$ depending on the precise oxygen isotopic ratio. This also appears to be consistent with optical depth estimates based on $^{12}$CO/$^{13}$CO ratios \citep{2004ApJ...611..835P}. 

Note, however, that in general we measure $r_{21}<1$. This departure is not due to subthermal excitation, or exceedingly cold gas temperatures. The former would predict extremely low HCN/CO line ratios, contrary to what is observed, while the latter would predict low excitation for the higher CO transitions, which is not observed either (\citealt{2003ApJ...586..891B}, \citealt{2011ApJ...735...19S},  \citealt{2015ApJ...801...63M}) Most likely the low ratios are due to temperature gradients, as discussed in \S\ref{interpreting_r21} \citep[see also ][]{1974ApJ...187L..67S,1984ApJ...287..153Y}.    
The existence of regions with measured $r_{21}<$1 is consistent with observations of other sources (e.g., \citealt{1992A&A...264..433B}, \citealt{2009AJ....137.4670L}, \citealt{2013AJ....146...19L}, \citealt{2017ApJS..233...22S}).  For example, resolved observations of giant molecular clouds in Orion by \citet{1994ApJ...425..641S} and \citet{2015ApJS..216...18N} also find ratios on the order of $\sim$0.6 in ``intermediate regions" -- i.e.~those regions not coincident with {\sc H\,ii} regions or ridges. Some of these values may be attributable to subthermal excitation or low temperatures, but because of how easy it is to excite the CO(2--1) transition it seems apparent that most of them must be due to temperature gradient effects.

The situation for the different measurements in the SW streamer and other outflow features, is similar. There the values of $r_{21}$, if anything, cluster closer to unity.
The most natural conclusion is that the majority of the molecular gas has conditions such that the CO emission is also optically thick throughout the SW streamer and other outflow features. It is possible to create a scenario where the emission is optically thin but the densities are finely tuned to not excite CO(2--1) efficiently. It seems, however, extremely unlikely that such a finely tuned situation would be attained throughout the outflowing gas. It is also possible to finely tune the temperature to create a similar effect, but that scenario is even more contrived because of the low energies of both transitions.

The optical depth of the CO emission has an impact on the mass estimate for the outflow features, and the precise molecular outflow rate of the source.

\subsection{Implications for the Molecular Outflow Rate}
 
\citet{2013Natur.499..450B} assumed optically-thin emission from warm gas to compute lower limits to the molecular mass and the mass loss rate due to the outflow in NGC\,253 of $6.6\times10^6$~M$_\odot$ and 9~M$_\odot$\,yr$^{-1}$ respectively.  

The results presented here, together with those from \citet{2017ApJ...835..265W}, suggest that the real masses and mass loss rates are larger, perhaps considerably larger.
Based on the detection of high-density tracers \citet{2017ApJ...835..265W} outlined a scenario where, although it is perhaps possible to excite the HCN emission in the ``minimum mass'' SW streamer, it would be much easier if its mass were a factor of $\sim5$ higher. The optically thin
CO-to-H$_{2}$ conversion factor employed in these calculations was $\alpha_{\rm CO}=0.34$~M$_{\odot}$\,(K\,km\,s$^{-1}$\,pc$^2)^{-1}$  \citep{2013Natur.499..450B}. This is a factor of $\sim13$ times lower than the CO-to-H$_{2}$ conversion factor for self-gravitating giant molecular clouds in the Milky Way disk and in the disks of other normal metallicity galaxies \citep{2013ARA&A..51..207B}. In the case of NGC\,253 and other starbursts the conversion factor for the gas producing the bulk of the CO luminosity is frequently lower, due to a combination of higher temperature and increased velocity dispersion in gas that is not self-gravitating. 

The best estimates for the starburst region of NGC\,253 constrain the CO-to-H$_2$ conversion factor there to be  
$\alpha_{\rm CO}=1-2$~M$_{\odot}$\,(K\,km\,s$^{-1}$\,pc$^2)^{-1}$ 
(\citealt{2011ApJ...735...19S}, \citealt{2015ApJ...801...25L}), 3 to 6 times larger than the optically thin value \citep{2013Natur.499..450B}. If we adopt this range of conversion factors  for all the molecular gas in the outflow, the implied molecular masses and mass loss rates would accordingly increase to ${\rm M_{H2}}\sim2-4\times10^7$~M$_\odot$ and ${\rm \dot{M}_{H2}}\sim25-50$~M$_\odot$\,yr$^{-1}$ respectively. The resulting mass-loading parameter for the starburst-driven molecular wind, $\eta={\rm \dot{M}_{H2}/SFR}$, would be in the range of 9 to 19. Needless to say, there are considerable uncertainties associated with these values beyond the adoption of a conversion factor. Among them, are the precise apportioning of CO emission to the outflow, the geometry of the outflow built into the corrections for projection, and the fraction of the outflowing gas that actually escapes the host galaxy.  

Observed trends between $\eta$ and outflow energetics are presented in \citet{2014A&A...562A..21C}.  The adjusted $\eta$ for NGC 253 does not set it apart from other starburst galaxies in that sample as an increased $\alpha_{CO}$ is not a trait intrinsic to the observations and would presumably apply to the entire sample.  Moreover, an $\eta$ of 10--20 (especially when a significant fraction of the outflowing material may be reaccreted onto the host galaxy) is consistent with the result of simulations (e.g., 
\citealt{2012MNRAS.421.3522H},
\citealt{2015MNRAS.454.2691M}).

\subsection{The Nature of the CO(2--1) Extended Component}\label{haze}

The low-level extended component seen in the CO(2--1) merits further discussion. We assessed whether this emission is real in \S\ref{extended_component_test}, and it appears that it is very difficult to explain it as an artifact of the calibration or the combination. Assuming that this feature is indeed real, we now consider its morphology and kinematics and discuss possible physical interpretations. 
    
Examination of the channel maps from $-90.5$ to 110.5 km\,s$^{-1}$ in Figure~\ref{channel_maps} shows that the extended emission component spans the region between the central starburst and the southeastern spiral arm. In the same channels, the extended component spatially overlaps with outflow features both above and below the midplane.  Furthermore, the extended component seems most prominent in the western half of NGC\,253, in particular to the southwest where the SW streamer and additional compact outflow features are located (Figure~\ref{spectra_map_LeroyMatch}).

\begin{figure*}
\includegraphics[width=180mm]{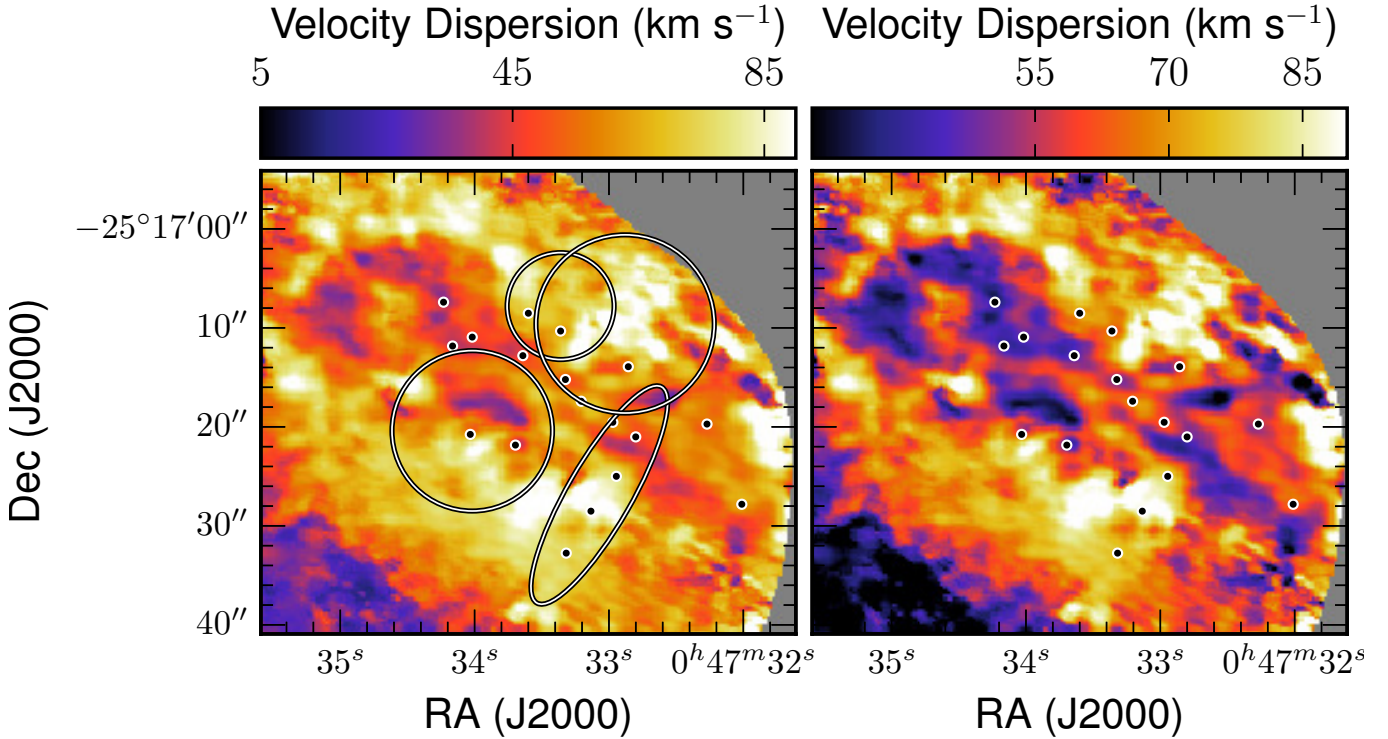}
\caption{\textit{CO(2--1) dispersion maps including all emission above a 5$\sigma$ threshold  (11 mJy beam$^{-1}$ in a 5 km s$^{-1}$ channel).  The two panels are the same, but with different boundaries on their color scales(left is 0--90 km s$^{-1}$ and right is 50--90 km s$^{-1}$). In both panels, emission extending above the midplane clearly has higher velocity dispersion than does the emission within the disk.  The right panel excludes most in-disk emission and shows hints of a biconical structure. Ellipses and marks are as in Figure~\ref{channel_maps_ratio_21}. For clarity, ellipses are only included in the left-most panel.} \label{dispersion_CO21_clipped_2panel}}
\end{figure*}

The velocity dispersion map in Figure~\ref{dispersion_CO21_clipped_2panel} shows higher dispersions in the emission above and below the central region, as high as $\sim$90 km s$^{-1}$ and more in some cases. This compares to the velocity dispersion in the starburst region, which is generally lower than 50 km s$^{-1}$.  
Interestingly, the higher velocity dispersion regions show evidence for a biconical structure, narrowing at the central starburst region and expanding above and below it.  This structure is best seen in the right panel of Figure~\ref{dispersion_CO21_clipped_2panel}, where the color bar has been adjusted to show only regions with velocity dispersion higher than 50~km\,s$^{-1}$. This biconical structure resembles the structure of the ionized wind in NGC\,253 \citep{2011MNRAS.414.3719W}. It is plausible that what we are seeing is a faint, extended component of CO-emitting molecular gas that completely surrounds the ionized cone. Enhancements in the CO emission such as the SW streamer at the other individually identified features may be due to individual clouds entrained in the outflow, that are found in the process of being disrupted and ejected. But the entire ionized outflow may be encased in a sheath of faint CO emission created by the ejection of molecular material from the largely molecular environment of the starburst region.

\begin{figure}
\includegraphics[width=80mm]{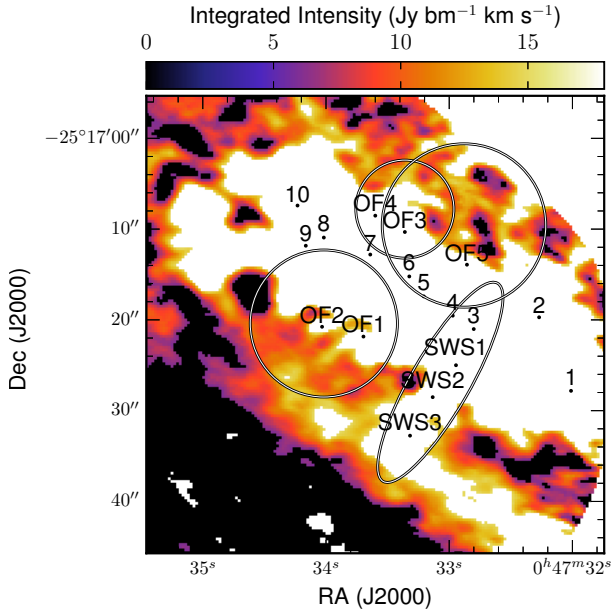}
\caption{\textit{A zeroth-moment map excluding as much main-disk and dense outflow emission as possible. The map is created via masking emission outside the range of 6--15 mJy beam$^{-1}$ km s$^{-1}$ in each 5 km s$^{-1}$ channel, as well as setting a velocity dispersion requirement of 50 km s$^{-1}$ prior to masking. Regions with emission below this range are shown in black and regions with emission above this range are shown in white.  Only the unmasked emission shown here is included in our estimate of the mass of the extended component.  Ellipses and regions are the same as those marked in Figure~\ref{spectra_map_LeroyMatch}.} \label{masked_moment}}
\end{figure}
 
\par
In an attempt to include only the extended component, we create a zeroth-moment map by including only emission from 6--15 mJy beam$^{-1}$ km s$^{-1}$ in each 5 km s$^{-1}$ channel (Figure~\ref{masked_moment}). A second criteria is that the included emission must also have a velocity dispersion higher than 50 km s$^{-1}$ (prior to making the 6--15 mJy beam$^{-1}$ km s$^{-1}$ constraint, when all emission above 2$\sigma$ is included in a second-moment map).  It may be that more emission is included in the extended component, but we set a strict threshold in order to avoid including the edges of distinct outflow features. We first determine that the diffuse, high-dispersion component accounts for $\sim$15$\%$ of the total emission in the central region and outflow (not including emission associated with the spiral arm visible to the southwest).  This fraction is consistent with the fraction of diffuse CO observed in other galaxies (e.g.\ \citealt{2013ApJ...779...43P}, \citealt{2013AJ....146..150C}).   The corresponding luminosity for the high velocity dispersion emission is 2.7$\times$10$^{4}$ K km s$^{-1}$ pc $^2$.  If we interpret this as likely optically thin CO(2--1) emission from a diffuse and warm molecular component, we can estimate its mass by applying a CO(1--0) optically-thin conversion factor of ${\alpha_{\rm CO}}=$0.34 M$_{\odot}$ (K km s$^{-1}$ pc $^2$)$^{-1}$, and assuming r$_{21}\sim$3--4. The resulting mass is highly uncertain, given the assumptions, but on the order of $\sim$2.6$\times$10$^{4}$ M$_{\odot}$.

We first estimate that this component extends approximately 20$\arcsec$ above the midplane on either side.  Its 

morphology and the $78^{\circ}$ inclination of the galaxy render  projection effects (for example, emission from the midplane of the NGC\,253 disk seen in projection and interpreted as emission extended above the disk) difficult to gauge. Since the galaxy is not edge-on, we must estimate the radial extent of the emission to gauge the effect of projection.  For this purpose, we assume that the high velocity dispersion emission extends $\sim$30$\arcsec$ (500 pc) in radius ($R$) on either side of the central region.   Thus, the minimum deprojected height assuming all emission is from the midplane is $h_{mid}=R\,\cos i$,  corresponding to $\sim100$ pc.   To obtain the maximum height, we ignore these projection effects, and instead directly convert the observed height $h_{obs}$ of 20$\arcsec$ to pc. This yields a maximum height of 340 pc.  The disk is clearly not infinitely thin, nor is the disk perfectly edge-on. Thus, the true height of the emission above the midplane is somewhere between 100 and 340 pc.  

This extent above the midplane is much less than that of the extended CO(2--1) component detected in M 82 by \citet{2015ApJ...814...83L}.  However, the maps of NGC 253 presented here cover a much smaller region than that observed in M 82 -- with the area mapped in M 82 being a factor of $\sim$30 larger.  Furthermore, there are clear indications in Figure~\ref{channel_maps} that the CO(2--1) extends well beyond the currently mapped region of NGC 253.    Thus, these observations do not preclude such an extended component in NGC 253.  The emission detected in NGC 253 may also be consistent with the aforementioned diffuse CO emission observed in normal galaxies (e.g., \citealt{2013ApJ...779...43P}, \citealt{2013AJ....146..150C}).

\section{Summary}\label{summary}

We present and analyze the $^{12}$CO(1--0) and $^{12}$CO(2--1) emission in the central regions and galactic wind of NGC\,253, yielding the following results:

\par 
1) The ratio of the brightness temperatures of CO(2--1) and CO(1--0), $r_{21}$, is close to unity for a majority of cases within both the disk and the outflow (the SW streamer, for example), indicating that the CO(1--0) emission is optically thick.  In fact most values of $r_{21}$ are under unity. We attribute this to temperature gradients in the optically thick emitting gas. The values of $r_{21}$ we measure are otherwise similar to those common in a range of environments such as galaxy disks.  It should be noted that $r_{21}$ is rather insensitive to environmental factors.

2) The fact that the bulk of the emission in the outflow has $r_{21}\lesssim1$ implies that mass estimates based on an optically thin CO-to-H$_2$ conversion factor will underestimate its true mass.
If, lacking a better constraint, we adopt the same conversion factor observed in the central region of the NGC\,253 starburst for all the emission associated with the molecular outflow, following the calculations of \citet{2013Natur.499..450B} the implied molecular mass outflow rate increases by factors of $3-6$ to 
${\rm \dot{M}_{H2}}\sim25-50$~M$_\odot$\,yr$^{-1}$. The resulting mass-loading parameter would be $\eta\sim9-19$.

3)  We report the tentative detection of a faint CO(2--1) extended emission component.  This low-level emission has very little impact on our measured $r_{21}$ values. The fraction of this emission with high velocity dispersion ($>50$~km\,s$^{-1}$) appears to be distributed on a biconical structure approximately coincident with the ionized outflow.

\section{Acknowledgments}\label{acknowledgments}

We thank the anonymous referee for exceptionally 
constructive suggestions leading to the improvement of this manuscript. This paper makes use of the following ALMA data: ADS/JAO.ALMA\#2011.1.00172.S, 2012.1.00108.S, and 2013.1.00191.S. ALMA is a partnership of ESO (representing its member states), NSF (USA) and NINS (Japan), together with NRC (Canada), NSC and ASIAA (Taiwan), and KASI (Republic of Korea), in cooperation with the Republic of Chile. The Joint ALMA Observatory is operated by ESO, AUI/NRAO and NAOJ. We thank the ALMA operators, staff, and helpdesk for making this work possible.  This research made use of APLpy, an open-source plotting package for Python \citep{2012ascl.soft08017R}, and Matplotlib \citep{Hunter:2007}.  LKZ greatly appreciates support from the Finnish Center for Astronomy with ESO (FINCA), comprised of the University of Turku, the University of Helsinki, Aalto University, and the University of Oulu, in conjunction with the European Southern Observatory.  The work of ADB is supported in part by the NSF under grant AST-1412419. JMDK gratefully acknowledges funding from the German Research Foundation (DFG) in the form of an Emmy Noether Research Group (grant number KR4801/1-1) and from the European Research Council (ERC) under the European Union's Horizon 2020 research and innovation programme via the ERC Starting Grant MUSTANG (grant agreement number 714907).  AKL is partially supported by the National Science Foundation under Grants No. 1615105, 1615109, and 1653300.  SV gratefully acknowledges NSF grant AST-1009583.  We thank the ALMA operators and staff, as well as the ALMA helpdesk for  diligent feedback and invaluable assistance in processing these data. We thank Nick Scoville for constructive discussion.

\bibliographystyle{apj}
\bibliography{ngc253_zschaechner_etal}

\end{document}